\documentclass{jfm}
\usepackage{graphicx}
\usepackage{epstopdf, epsfig}
\usepackage{bm,amsfonts}

\shorttitle{Theory and Simulation of Mesoscale Electrokinetic Fluctuations}
\shortauthor{M. Deng, F. Tushar, L. Bravo, A. Ghoshal, G. Karniadakis and Z. Li}

\title{Theory and simulation of electrokinetic fluctuations in electrolyte solutions at the mesoscale}

\author{Mingge Deng\aff{1},
  Faisal Tushar\aff{2},
  Luis Bravo\aff{3},
  Anindya Ghoshal\aff{3},
  George Karniadakis\aff{1},
 \and Zhen Li\aff{2}
 \corresp{\email{zli7@clemson.edu}}
 }

\affiliation{
\aff{1} Division of Applied Mathematics, Brown University, Providence, RI 02912, USA
\aff{2} Department of Mechanical Engineering, Clemson University, Clemson, SC 29634, USA
\aff{3} US Army Research Laboratory, Aberdeen Proving Ground, MD 21005, USA
}

\begin{document}

\maketitle

\begin{abstract}
Electrolyte solutions play an important role in energy storage devices, whose performance highly relies on the electrokinetic processes at sub-micron scales.\ Although fluctuations and stochastic features become more critical at small scales, the long-range Coulomb interactions pose a particular challenge for both theoretical analysis and simulation of fluid systems with fluctuating hydrodynamic and electrostatic interactions. Here, we present a theoretical framework based on the Landau-Lifshitz theory to derive closed-form expressions of fluctuation correlations in electrolyte solutions, indicating significantly different decorrelation processes of ionic concentration fluctuations from hydrodynamic fluctuations, which provides insights for understanding transport phenomena of coupled fluctuating hydrodynamics and electrokinetics. Furthermore, we simulate fluctuating electrokinetic systems using both molecular dynamics (MD) with explicit ions and mesoscopic charged dissipative particle dynamics (cDPD) with semi-implicit ions, from which we identify that the spatial probability density functions of local charge density follow Gamma distribution at sub-nanometer scale (i.e., $0.3~{\rm nm}$) and converge to Gaussian distribution above nanometer scales (i.e., $1.55~{\rm nm}$), indicating the existence of a lower limit of length scale for mesoscale models using Gaussian fluctuations. The temporal correlation functions of both hydrodynamic and electrokinetic fluctuations are computed from all-atom MD and mesoscale cDPD simulations, showing a good agreement with the theoretical predictions based on the linearized fluctuating hydrodynamics theory.
\end{abstract}


\section{Introduction}\label{intr}
Electrolyte solutions, consisting of a polarized solvent and ionic species, are extremely important in a wide range of fields including chemical physics, biology, and geochemistry~\citep{2013Guglielmi,2020Molina}. They are also of interest in engineering physics, such as aggregation and dispersion of charged liquid particles~\citep{2020Boutsikakis,2020Nieto} present in aerospace systems.
Despite a long history of electrolyte solution studies, there are still important open questions associated with fluctuations and correlations of electrolyte bulk solutions~\citep{2019Donev}.
At the mesoscale (i.e., nanometer to micrometer length scales), thermal energies of electrolyte solutions are of the same magnitude as the characteristic energies of hydrodynamics and electrokinetics.
Therefore, thermally induced fluctuations can play an important role in both equilibrium and non-equilibrium electrokinetic phenomena at the mesoscale~\citep{2021Ladiges}. Quantifying the impact of fluctuations on mesoscale fluid systems is critical to understanding the large-scale dynamics of complex fluid systems designed from the nano/mesoscale in a bottom-up approach. 

The essential feature of an electrolyte solution is that charged species interact with each another with long-range Coulomb forces, leading to a system whose properties are significantly different from those of electrically neutral solutions~\citep{2020Klymko}.
At the continuum scale level, fluctuating hydrodynamic and electrokinetic theories have been used to describe thermally induced fluctuations  through random tensor/flux terms in the governing equations, formulated properly to satisfy the fluctuation-dissipation theorem (FDT)~\citep{Landau1959, Ortiz2006}.
At the atomic scale, molecular dynamics (MD) simulations with explicit ions have also been used to study electrolyte solutions. However, the computational cost of MD simulations further prevents their use at large length- and time-scales~\citep{Chen2007,Joung2013,Yoshida2014}. 
Mesoscale approaches smoothly bridge the gap between continuum and atomistic descriptions and provide the possibility to consistently integrate fluctuations into macroscopic field variables. 
To this end, we developed a charged dissipative particle dynamics (cDPD) model~\citep{2016Deng} to simulate mesoscale electrokinetic phenomena with fluctuations. 

In the present work, we consider a continuum formulation and derive the coupled fluctuating hydrodynamic and electrokinetic equations. Based on the linear response assumption~\citep{1982Kubo}, we will linearize the fluctuating hydrodynamic and electrokinetic equations and derive analytical closed-form expressions for current correlation functions of electrolyte bulk solutions in Fourier space. Then, we will perform both all-atom MD simulations with explicit ions and mesoscale cDPD simulations with semi-implicit ions to compute spatial and temporal correlations of hydrodynamic and electrostatic fluctuations directly from particle trajectories. These simulation results will be used to validate the closed-form expressions for correlation functions derived from the linearized fluctuating hydrodynamics theory by comparing simulation results against theoretical predictions.

The remainder of this paper is organized as follows:  In section~\ref{sec:theory} we introduce the continuum formulation for fluctuating hydrodynamics and electrokinetics, and also the derivations of analytical solutions for current correlation functions using perturbation theory. In section~\ref{sec:results} we describe the details for performing all-atom MD simulations and mesoscopic cDPD simulations, and we also present the simulation results comparing against the theoretical predictions. Finally, we conclude with a brief summary and discussion in Section~\ref{sec:summary}.

\section{Continuum theory}\label{sec:theory}
\subsection{Fluctuating hydrodynamics and electrokinetics}
We consider a mesoscale system of electrolyte solution at thermal equilibrium in a periodic domain $\mathbf{\Omega}$ of fixed volume. The system contains $S$ types of ionic species with concentration $c_\alpha(\mathbf{r},t)$ and charge valency $z_\alpha$ for the $\alpha$-th ionic specie.
Let $e$ be the elementary charge, then a global constraint is imposed by the charge neutrality condition
\begin{equation}
\sum_{\alpha=1}^S \int_{\mathbf{\Omega}} e z_\alpha c_\alpha(\mathbf{r},t)d\mathbf{r} = 0. \nonumber
\end{equation}
The solvent molecules are represented implicitly through their electrostatic and thermodynamic properties such as dielectric constant $\epsilon$, bulk viscosity $\zeta$, and shear viscosity $\eta$.
From the continuum perspective, this system can be described by equations of classical fluctuating hydrodynamics with an additional electrostatic force
\begin{subeqnarray}\label{eq:gov}
	&\frac{\partial \rho}{\partial t} + \nabla \cdot \mathbf{g} = 0,  \\
	&\frac{\partial \mathbf{g}}{\partial t} + \nabla \cdot (\mathbf{g}\mathbf{v}) = - \nabla p + \eta\nabla^2\mathbf{v} + (\frac{\eta}{3} + \zeta) \nabla(\nabla \cdot \mathbf{v}) 
    + \rho_e \mathbf{E} + \nabla \cdot \delta \mathbf{\bm{\Pi}},
\end{subeqnarray}
for velocity $\mathbf{v}(\mathbf{r}, t)$, pressure $p(\mathbf{r},t)$, mass density $\rho(\mathbf{r},t)$, momentum density $\mathbf{g}(\mathbf{r},t) = \rho(\mathbf{r},t) \mathbf{v}(\mathbf{r},t)$,
and electric field $\mathbf{E}(\mathbf{r},t)$. The local charge density is given by $\rho_e(\mathbf{r},t) = \sum_\alpha^S e z_\alpha c_\alpha(\mathbf{r},t)$.
The random stress tensor $\delta \bm{\Pi}$ is a matrix of Gaussian-distributed random variables with zero means and variances given by the fluctuation-dissipation theorem (FDT)~\citep{Ortiz2006}
\begin{equation}\label{eq:FDT}
	\langle \delta \mathbf{\Pi}_{ij}(\mathbf{r},t) \delta \mathbf{\Pi}_{kl}(\mathbf{r}',t')\rangle = 2 k_BT \mathbf{C}_{ijkl}
	\delta(\mathbf{r}-\mathbf{r}')\delta(t-t'),
\end{equation}
where $\mathbf{C}_{ijkl}=\eta(\delta_{ik}\delta_{jl} + \delta_{il}\delta_{jk})+(\zeta-{2\eta}/{3}) \delta_{ij}\delta_{kl}$ is a rank-4 tensor.
The fluctuating hydrodynamic equations are closed with the equation of state, e.g., $c_s^2 = (\partial p/\partial \rho)_T$ with $c_s$ being the isothermal sound speed.

The Ginzburg-Landau free energy functional for an electrolyte solution is~\citep{1950Ginzburg}
\begin{equation}\label{eq:free_energy}
	G = \int_{\mathbf{\Omega}} d\mathbf{r} \bigg\{k_BT \sum_\alpha c_\alpha \ln c_\alpha
	+ \sum_\alpha e z_\alpha c_\alpha \phi - \frac{\epsilon}{2}(\nabla\phi)^2\bigg\} , \nonumber
\end{equation}
where $k_BT$ is the thermal energy and $\phi$ the electrostatic potential.
Variation of the free energy with respect to the electrostatic potential yields the Poisson equation
\begin{equation}\label{eq:poisson}
	-\nabla \cdot (\epsilon(\mathbf{r}) \nabla \phi ) = \sum_\alpha^S e z_\alpha c_\alpha(\mathbf{r},t),
\end{equation}
by setting $\delta G/\delta \phi = 0$.
Similarly, the electrochemical potential of the $\alpha$-th ionic species can be derived by variation with respect to ionic concentration
\begin{equation}\label{eq:chemical_potential}
	\mu_\alpha  = \frac{\delta G}{\delta c_\alpha} = k_BT \ln c_\alpha + z_\alpha e\phi. \nonumber
\end{equation}
The transport and dissipation of ionic species are driven by the fluid velocity, electrochemical potential, and thermal fluctuations, which can be described in terms of the ionic concentration flux $\mathbf{J}(\mathbf{r},t)$,
\begin{equation}
	\frac{\partial c_\alpha(\mathbf{r},t)}{\partial t} + \mathbf{v}(\mathbf{r},t) \cdot \nabla c_\alpha(\mathbf{r},t) =
	-\nabla\cdot(\mathbf{J}_{\alpha}(\mathbf{r},t) + \delta \mathbf{J}_{\alpha}(\mathbf{r},t) ), \nonumber
\end{equation}
where $\mathbf{J}_{\alpha}(\mathbf{r},t)$ is the dissipative flux and $\delta \mathbf{J}_{\alpha}(\mathbf{r},t)$ is the random flux when the system is near thermodynamic equilibrium.
The diffusion flux can be written as
$\mathbf{J}_\alpha = - \sum_\beta^S M_{\alpha \beta}\nabla \mu_\beta (\mathbf{r},t)$,
in which $\nabla \mu$ is the thermodynamic force for diffusion flux and $M_{\alpha \beta}$ are the Onsager coefficients related to macroscopic ionic diffusion coefficients.
In general, $M_{\alpha \beta} = M_{\beta \alpha} \neq 0$ as implied by reversal invariance~\citep{1931Onsager1,1931Onsager2}.
The off-diagonal terms $M_{\alpha \neq \beta}$ describe mutual-diffusion and are assumed to be relatively small compared to the diagonal self-diffusion terms.
For example, according to experimental data~\citep{1967Chapman}, the self-diffusion coefficients of cation and anion in a $1~{\rm M}$ NaCl solution are $1.16\times 10^{-9} {\rm m^2/s}$ and $1.99\times 10^{-9} {\rm m^2/s}$, respectively,
while the mutual-diffusion coefficient is one order of magnitude smaller: $1.3\times 10^{-10} {\rm m^2/s}$.
Therefore, we assume that the mutual diffusion terms can be ignored in the present work for simplicity. In numerical experiments, this assumption is checked by computing the self- and mutual-diffusion coefficients from MD data using the Green-Kubo relations~\citep{1996Zhou,2004Wheeler}, which is presented in section~\ref{sec:results}.

By substituting the electrochemical potential expression into the diffusion flux, we obtain
\begin{equation}
	\mathbf{J}_\alpha = - M_{\alpha}\nabla \mu_\alpha (\mathbf{r},t) = - \frac{M_{\alpha} k_BT}{c_\alpha} \nabla c_\alpha - M_{\alpha}z_\alpha e\nabla \phi .  \nonumber
\end{equation}
In practice, it is more convenient to use the macroscopic diffusion coefficients $D_\alpha = {M_{\alpha} k_BT}/{c_\alpha}$ instead of the phenomenological coefficients $M_{\alpha}$. Therefore, the ionic concentration transport equation can be rewritten as
\begin{subeqnarray}\label{eq:pnp}
	\frac{\partial c_\alpha(\mathbf{r},t)}{\partial t} + \mathbf{v}(\mathbf{r},t) \cdot \nabla c_\alpha(\mathbf{r},t) 
	= \nabla \cdot \Big(D_\alpha \nabla c_\alpha + \frac{D_\alpha e z_\alpha c_\alpha}{k_BT}\nabla \phi +\delta \mathbf{J}_{\alpha}(\mathbf{r},t) \Big).
\end{subeqnarray}
For a system near thermodynamic equilibrium, the random fluxes can be modeled as Gaussian random vectors with zero means and variance given by the generalized FDT as
\begin{subeqnarray}\label{eq:FDT_pnp}
\langle \delta \mathbf{J}_{\alpha} (\mathbf{r},t) \cdot \delta\mathbf{J}_{\beta}(\mathbf{r}',t')\rangle 
&&= 2k_BT M_{\alpha \beta} \delta (\mathbf{r} -\mathbf{r}') \delta(t-t')\nonumber\\
&&= 2 D_{\alpha} c_{\alpha}\delta (\mathbf{r}-\mathbf{r}') \delta(t-t').
\end{subeqnarray}
The coupled Eqs.~(\ref{eq:gov})-(\ref{eq:FDT_pnp}) form the fluctuating hydrodynamic and electrokinetic equations.

\subsection{Linearized theory of electrolyte solution}
The stochastic partial differential equations (SPDEs) given by Eqs.~(\ref{eq:gov})-(\ref{eq:FDT_pnp}) could be solved through numerical discretization, with the FDT satisfied on the discrete level following the GENERIC framework~\citep{1997Grmela,1997Ottinger}.
In general, it is very challenging to obtain an analytical solution of such sPDEs.
However, if the local hydrodynamic and electrokinetic fluctuations are sufficiently small, linear response theory can be applied to derive linearized equations to describe the relaxation process of electrolyte solution towards equilibrium. In the present work, we focus on the linearized equations for electrolyte solutions and their closed-form solutions.

The equilibrium state of a bulk electrolyte solution is characterized by its mean field properties, i.e., constant mass density $\rho_0$, constant pressure $p_0$, constant bulk ionic concentration $c_{\alpha 0}$, zero momentum field $\mathbf{g}_0 = 0$, and zero electrostatic potential field $\phi_0 = 0$.
The local fluctuating hydrodynamic field can be expressed as the perturbation around the mean field state:
\begin{subeqnarray}
		&&\rho(\mathbf{r},t) = \rho_0 + \delta \rho(\mathbf{r},t), \\
		&&p(\mathbf{r},t) = p_0 + \delta p(\mathbf{r},t),\\
		&&\mathbf{g}(\mathbf{r},t) = \delta \mathbf{g} (\mathbf{r},t),
\end{subeqnarray}
where the local perturbation of pressure can be related to density fluctuations via the equation of state $\delta p = c_s^2 \delta \rho$ under isothermal conditions.
Also, the local fluctuating electrostatic field can be decomposed as
\begin{subeqnarray}
		c_\alpha(\mathbf{r},t) &&= c_{\alpha 0} + \delta c_\alpha(\mathbf{r},t), \\
		\phi(\mathbf{r},t) &&= \delta \phi(\mathbf{r},t),\\
		\rho_e(\mathbf{r},t) &&= \delta \rho_e (\mathbf{r},t),
\end{subeqnarray}
in which $\delta\rho_e(\mathbf{r},t) = \sum_\alpha^S e z_\alpha \delta c_\alpha(\mathbf{r},t)$ when the global electro-neutrality condition $\sum_\alpha^S e z_\alpha c_{\alpha 0} = 0$ is imposed.
The fluctuating electrostatic potentials and charge densities are related via the Poisson equation.

For small fluctuations of electrostatic field, the linearized fluctuating hydrodynamic and electrokinetic equations can be written as
\begin{subeqnarray}
		\frac{\partial (\delta \rho)}{\partial t} &&= -\nabla \cdot \delta \mathbf{g}, \\
		\frac{\partial (\delta \mathbf{g})}{\partial t} &&= -{c_s^2}\nabla \delta \rho + \eta \nabla^2 (\delta \mathbf{v}) + (\frac{\eta}{3}+\zeta) \nabla (\nabla \cdot\delta \mathbf{v} ),\\
		\frac{\partial (\delta c_\alpha)}{\partial t} &&= \Big(D_{\alpha} \nabla^2 (\delta
	c_{\alpha}) + D_{\alpha}\frac{e z_{\alpha} c_{\alpha 0}}{k_BT} \nabla^2 \delta \phi \Big), \\
		-\nabla \cdot (\epsilon \nabla \delta \phi) &&= \sum_\alpha^S e z_\alpha \delta c_\alpha(\mathbf{r},t),
\label{ns_pnp_lin}
\end{subeqnarray}
where only first-order perturbation terms are kept in these linearized equations.
It is important to note that the electrostatic force term $\delta \rho_e \nabla \delta \phi$ in the momentum equation and the convection term $\delta \mathbf{v} \cdot \nabla \delta c_\alpha$ in the transport equation are high-order perturbation terms and thus assumed to be negligible in the equations above. Therefore, the linearized hydrodynamic and electrokinetic equations become explicitly decoupled for fluid systems in thermodynamic equilibrium.

The linearized fluctuating hydrodynamics equations given by Eq.~(\ref{ns_pnp_lin}) can be transformed into $k$-space by a spatial Fourier transform and then solved in the $k$-space. Appendix~\ref{app:A} describes the detailed derivation of mass-momentum correlations by solving Eq.~(\ref{ns_pnp_lin}). 
The normalized temporal correlation functions (TCF) of the mass-momentum fluctuations in the $k$-space are given by
\begin{subeqnarray}\label{eq:ns_correlation}
		\frac{\langle \hat{\rho}(\mathbf{k},t)\hat{\rho}(\mathbf{k},0)\rangle}{\langle
      \hat{\rho}(\mathbf{k},0)\hat{\rho}(\mathbf{k},0)\rangle} &&= \exp(-\Gamma_T k^2 t) \cos(c_s k t), \\
		\frac{\langle \hat{\mathbf{g}}_\parallel(\mathbf{k},t)\hat{\mathbf{g}}_\parallel(\mathbf{k},0)\rangle}{\langle
      \hat{\mathbf{g}}_\parallel (\mathbf{k},0)\hat{\mathbf{g}}_\parallel(\mathbf{k},0)\rangle} &&= \exp(-\Gamma_T k^2 t) \cos(c_s k t), \\
		\frac{\langle
      \hat{\mathbf{g}}_\perp(\mathbf{k},t)\hat{\mathbf{g}}_\perp(\mathbf{k},0)\rangle}{\langle
        \hat{\mathbf{g}}_\perp (\mathbf{k},0)\hat{\mathbf{g}}_\perp(\mathbf{k},0)\rangle} &&= \exp(-\nu k^2 t), \\
		\frac{\langle \hat{\rho}(\mathbf{k},t) i
      \hat{\mathbf{g}}_\parallel(\mathbf{k},0)\rangle}{\langle \hat{\rho}
        (\mathbf{k},0) i \hat{\mathbf{g}}_\parallel(\mathbf{k},0)\rangle} &&= \exp(-\Gamma_T k^2 t) \sin(c_s k t),
\end{subeqnarray}
where the symbol $\hat{~}$ indicates Fourier components, and the wave vector $\mathbf{k} = (k, 0, 0)$ is defined along the (arbitrary) $x$-direction, $\nu = \eta/\rho$ is the kinematic viscosity, $\Gamma_T=2\nu/3+\zeta/2\rho$ is the sound absorption coefficient, and $c_s$ is the isothermal sound speed. Also,
$\mathbf{g}_\parallel$ represents longitudinal momentum (sound mode) parallel to the wave vector $\mathbf{k}$, and $\mathbf{g}_\perp$ represents the transverse component (shear mode) perpendicular to the wave vector $\mathbf{k}$.

For simplicity in deriving closed-form solutions, we only consider two types of ionic species: the cation (denoted by $p$) and anion (denoted by $n$).
By substitution of the Poisson equation into the linearized ionic transport equation, we obtain
\begin{subeqnarray}
		\frac{\partial(\delta c_p)}{\partial t} &= D_p \Big(\nabla^2 (\delta c_p) + \kappa_p^2 \delta c_p - \left| \frac{z_n}{z_p} \right| \kappa_p^2 \delta c_n \Big), \\
		\frac{\partial(\delta c_n)}{\partial t} &= D_n \Big(\nabla^2 (\delta c_n) + \kappa_n^2 \delta c_n - \left| \frac{z_p}{z_n} \right| \kappa_n^2 \delta c_p \Big),
	\label{pnp_lin}
\end{subeqnarray}
where $\kappa_p^2$ and $\kappa_n^2$ are defined as
\begin{subeqnarray}
		\kappa_p^2 &\equiv \frac{D_p {z_p}^2 c_{p 0} e^2}{k_BT \epsilon}, \quad\quad
		\kappa_n^2 &\equiv \frac{D_n {z_n}^2 c_{n 0} e^2}{k_BT \epsilon}
\end{subeqnarray}
with units of inverse-squared distance.
For a periodic system, we can expand the solution in Fourier modes:  $\delta c(\mathbf{r},t) = \sum_{\mathbf{k}}\delta \hat{c}(\mathbf{k},t) e^{i\mathbf{k}\cdot \mathbf{r}}$.
The spatial Fourier transformation of Eq.~(\ref{pnp_lin}) gives two coupled ordinary differential equations
\begin{subeqnarray}
		\frac{\partial \delta \hat{c}_p(\mathbf{k},t)}{\partial t} &= D_p\left[(\kappa_p^2-k^2) \delta \hat{c}_p(\mathbf{k},t) - \left| \frac{z_n}{z_p} \right| \kappa_p^2 \delta \hat{c}_n(\mathbf{k},t)\right], \\
		\frac{\partial \delta \hat{c}_n(\mathbf{k},t)}{\partial t} &= D_n\left[(\kappa_n^2-k^2) \delta \hat{c}_n(\mathbf{k},t) - \left| \frac{z_p}{z_n} \right| \kappa_n^2 \delta \hat{c}_p(\mathbf{k},t)\right].\nonumber
	\label{pnp_lin_k}
\end{subeqnarray}
The above equations can be written in matrix form as
$\displaystyle {\partial \mathbf{u}(\mathbf{k},t)}/{\partial t} + \mathcal{L}\mathbf{u} = 0$,
with the vector $\mathbf{u} = (\delta \hat{c}_p, \delta \hat{c}_n)^T$, and $\mathcal{L}$ the $2\times2$ matrix defined as
\begin{equation}
	\mathcal{L} = \left[
	\begin{array}{cc}
		(k^2 - \kappa_p^2)D_p & |\frac{z_n}{z_p}|\kappa_p^2 D_p \\
		|\frac{z_p}{z_n}|\kappa_n^2 D_n & (k^2 - \kappa_n^2) D_n \\
	\end{array}
	\right].
\end{equation}
The matrix $\mathcal{L}$ can be further split as $\mathcal{L} = -\mathcal{L}_0 + k^2 \mathcal{L}_1$, where
\begin{eqnarray}
		\mathcal{L}_0 &&= \left[ \begin{array}{cc}
			\kappa_p^2 D_p & -|\frac{z_n}{z_p}|\kappa_p^2 D_p \\
			-|\frac{z_p}{z_n}|\kappa_n^2 D_n & \kappa_n^2 D_n \\
		\end{array} \right] \\
		\mathcal{L}_1 &&= \left[ \begin{array}{cc}
			D_p & 0 \\
			0 & D_n \\
		\end{array} \right] ,
\end{eqnarray}
These equations can be solved by a linear combination of the eigenvectors $\xi^{(i)}(\mathbf{k})$, which satisfy the eigenvalue equation
\begin{equation}\label{eq:eigen}
\big[-\mathcal{L}_0 + k^2 \mathcal{L}_1 \big]\xi^{(i)}(\mathbf{k}) = \lambda_i\xi^{(i)}(\mathbf{k}).
\end{equation}
The conditions $\kappa_{p, n}^2 \gg k^2$ hold in the continuum limit, when either the domain size $L$ or the charge concentration $z_\alpha^2 c_{\alpha 0}$ are sufficiently large.
In this limit, the eigenvalue equation can be solved perturbatively by expanding $\xi^{(i)}$ and $\lambda^{(i)}$ in powers of $k^2$:
\begin{subeqnarray}\label{eq:expansion}
		\xi^{(i)} &= \xi_0^{(i)} + k^2 \xi_1^{(i)} + \cdots ,\\
		\lambda^{(i)} &= \lambda_0^{(i)} + k^2 \lambda_1^{(i)} + \cdots.
\end{subeqnarray}
Substitution of Eq.~(\ref{eq:expansion}) into the eigenvalue equation Eq.~(\ref{eq:eigen}) gives the zero- and second-order perturbation theory equations:
\begin{subeqnarray}
		\big(\mathcal{L}_0 - \lambda_0^{(i)}\mathbf{I}\big)\xi_0^{(i)} &&= 0,\\
		\big(\mathcal{L}_0 - \lambda_0^{(i)}\mathbf{I}\big)\xi_1^{(i)} &&= (\mathcal{L}_1 + \lambda_1^{(i)}\mathbf{I}) \xi_0^{(i)},
\end{subeqnarray}
where $\mathbf{I}$ denotes the identity matrix.
The solution to the order $\mathcal{O}(k^2)$ is given by two real negative roots approximated as $\lambda_1 \approx -\left((k^2+\kappa_p^2)D_p + (k^2+\kappa_n^2)D_n\right)+k^2 D_s(k)$ (fast decay) and $\lambda_2 \approx -k^2 D_s(k)$ (slow decay).
The collective diffusion coefficient of cation and anion pairs is defined as
\begin{equation}
	D_s(k) = \frac{(k^2 + \kappa_p^2 + \kappa_n^2)D_p D_n}{(k^2+\kappa_p^2)D_p + (k^2+\kappa_n^2)D_n},
\end{equation}
which has the same order as the diffusion coefficients $D_{\alpha}$.
The solutions of Eq.~(\ref{pnp_lin_k}) with initial conditions $\delta \hat{c}_p(\mathbf{k},0)$ and $\delta \hat{c}_n(\mathbf{k},0)$ can be written as
\begin{subeqnarray}
		\delta \hat{c}_p(\mathbf{k},t) &= A_{1} \delta \hat{c}_p(\mathbf{k}, 0) + A_{2} \delta \hat{c}_n(\mathbf{k}, 0), \\
		\delta \hat{c}_n(\mathbf{k},t) &= A_{3} \delta \hat{c}_n(\mathbf{k}, 0) + A_{4} \delta \hat{c}_p(\mathbf{k}, 0),
\end{subeqnarray}
with the coefficients
\begin{subeqnarray}
    &&A_{1} = \alpha_p e^{\lambda_1 t} + \alpha_n e^{\lambda_2 t},\\
    &&A_{2} = \beta_p(e^{\lambda_2 t} - e^{\lambda_1 t}),\\
    &&A_{3} = \alpha_n e^{\lambda_1 t} + \alpha_p e^{\lambda_2 t},\\
    &&A_{4} = \beta_n(e^{\lambda_2 t} - e^{\lambda_1 t}),
\end{subeqnarray}
where the dimensionless parameters are defined as
\begin{subeqnarray}
		\alpha_p &&= \frac{(k^2 + \kappa_p^2)D_p - k^2 D_s(k)}{(k^2 + \kappa_p^2)D_p + (k^2 +\kappa_n^2)D_n - 2 k^2 D_s(k)}, \\
		\alpha_n &&= \frac{(k^2 + \kappa_n^2)D_n - k^2 D_s(k)}{(k^2 + \kappa_p^2)D_p + (k^2 +\kappa_n^2)D_n - 2 k^2 D_s(k)}, \\
		\beta_p &&= \frac{|\frac{z_n}{z_p}|\kappa_p^2 D_p}{(k^2 + \kappa_p^2)D_p + (k^2 + \kappa_n^2)D_n - 2 k^2 D_s(k)}, \\
		\beta_n &&= \frac{|\frac{z_p}{z_n}|\kappa_n^2 D_n}{(k^2 + \kappa_p^2)D_p + (k^2 + \kappa_n^2)D_n - 2 k^2 D_s(k)},
\end{subeqnarray}
such that $\alpha_p + \alpha_n = 1$.

The fluctuations modeled by the above equations are transported and dissipated in time, and this can be shown via the time correlation of the fluctuating concentration field with different Fourier modes.
The temporal correlations between species $\alpha$ and $\beta$ in the fluctuating concentration field are given by
\begin{equation}
	\Psi_{\alpha \beta} = \frac{\langle \delta \hat{c}_\alpha(\mathbf{k},t) \delta \hat{c}_\beta^{*}(\mathbf{k},0)\rangle}{\langle \delta \hat{c}_\alpha(\mathbf{k},0)\delta \hat{c}_\beta^{*}(\mathbf{k},0)\rangle},
\end{equation}
where the symbol $^*$ denotes the complex conjugate.
The temporal correlation function for specific ionic species can then be expressed as
\begin{subeqnarray}\label{eq:TCF}
		\Psi_{p p} &&= \alpha_p e^{\lambda_1 t} + \alpha_n e^{\lambda_2 t} + \beta_p S_{n p} (e^{\lambda_2 t} - e^{\lambda_1 t}), \\
		\Psi_{n n} &&= \alpha_n e^{\lambda_1 t} + \alpha_p e^{\lambda_2 t} + \beta_n S_{p n} (e^{\lambda_2 t} - e^{\lambda_1 t}), \\
		\Psi_{p n} &&= \alpha_p e^{\lambda_1 t} + \alpha_n e^{\lambda_2 t} + \frac{\beta_p}{S_{p n}} (e^{\lambda_2 t} - e^{\lambda_1 t}), \\
		\Psi_{n p} &&= \alpha_n e^{\lambda_1 t} + \alpha_p e^{\lambda_2 t} + \frac{\beta_n}{S_{n p}} (e^{\lambda_2 t} - e^{\lambda_1 t}),
\end{subeqnarray}
in which $S_{p n}$ and $S_{n p}$ are static structure factor-like terms for a wave vector $\mathbf{k}$ defined as
\begin{subeqnarray}
		S_{p n}(\mathbf{k}) &= \frac{\langle \delta \hat{c}_p(\mathbf{k},0) \delta  \hat{c}_n^{*}(\mathbf{k},0)\rangle}{\langle\delta\hat{c}_n(\mathbf{k},0) \delta \hat{c}_n^{*}(\mathbf{k},0)\rangle},\\
		S_{n p}(\mathbf{k}) &= \frac{\langle \delta \hat{c}_n(\mathbf{k},0) \delta \hat{c}_p{*}(\mathbf{k},0)\rangle}{\langle\delta\hat{c}_p(\mathbf{k},0) \delta \hat{c}_p{*}(\mathbf{k},0)\rangle}.
\end{subeqnarray}
In this linearized theory, the covariance between the initial ionic concentration fluctuations of cation and anion are nonzero due to the charge neutrality constraint.
The structure factor terms $S_{p n}$ and $S_{n p}$ can be obtained from experiment or more detailed simulations.
The long time-behavior of both temporal auto-correlations and cross-correlations are dominated by slowly decaying terms that behave asymptotically as $\exp(\lambda_2 t)$, where $\lambda_2$ is one of the two (negative) eigenvalues.

\section{Numerical Results} \label{sec:results}
\begin{figure}
  \centering
  \includegraphics[width = 0.459\textwidth]{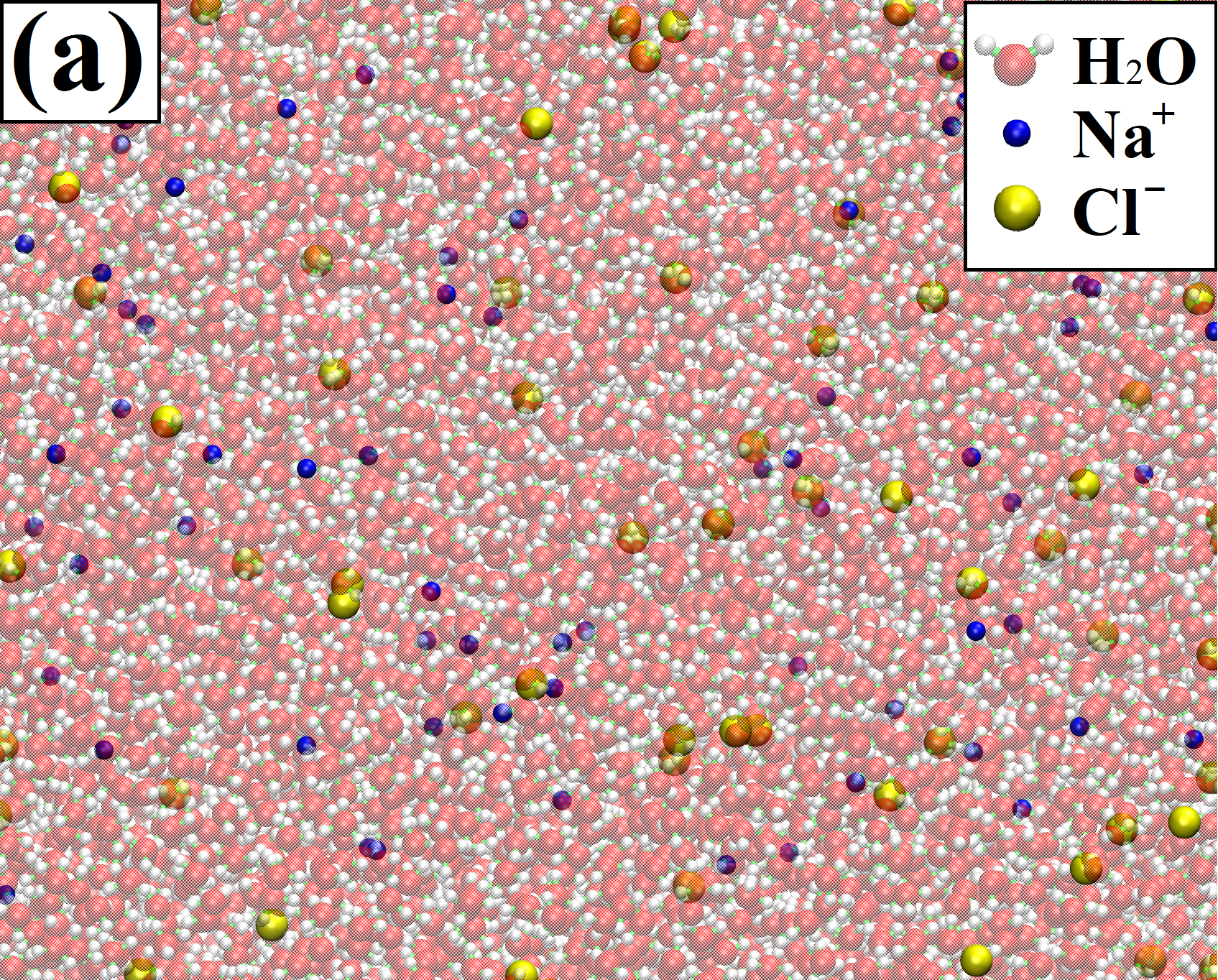}~~
  \includegraphics[width = 0.387\textwidth]{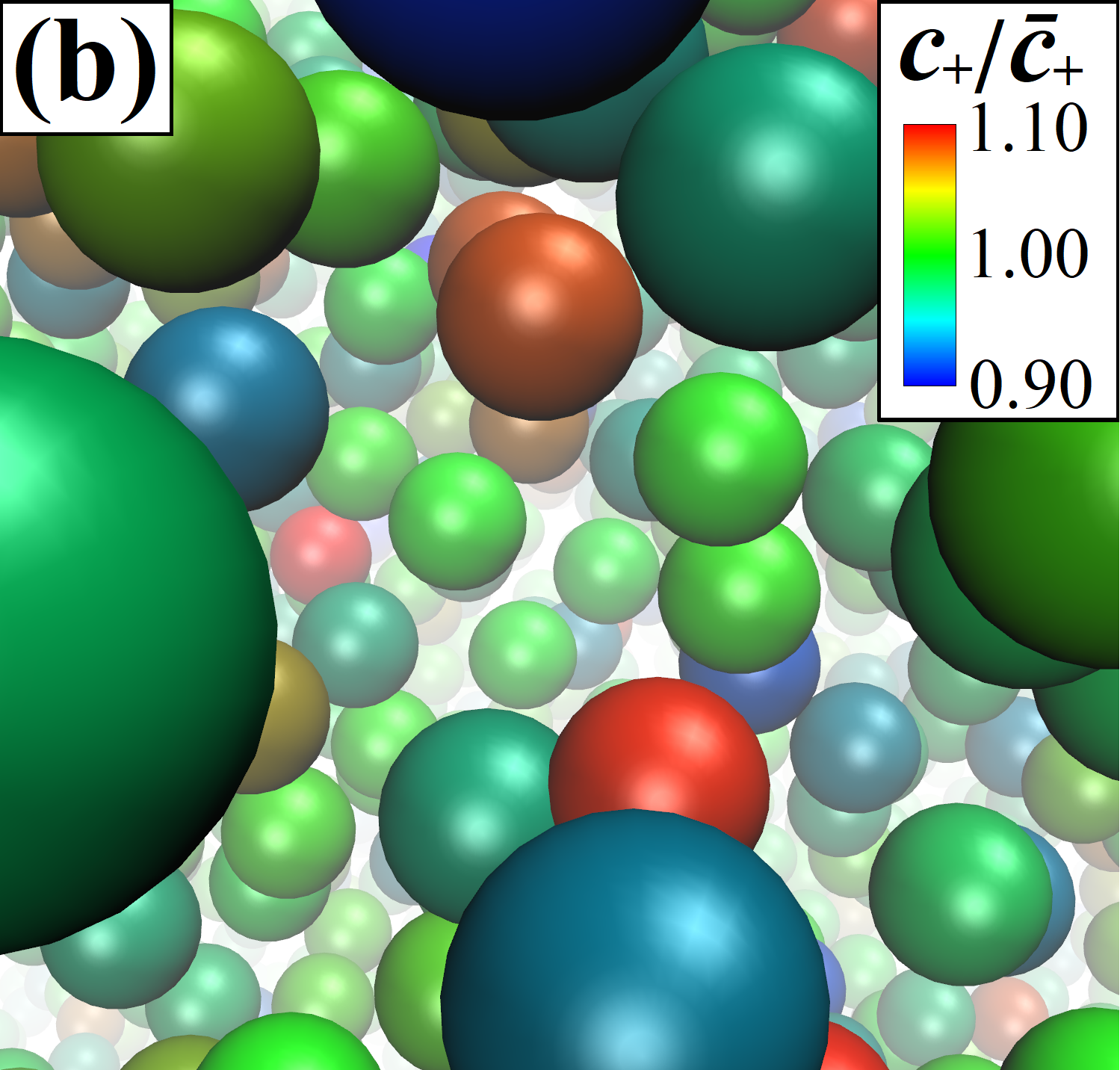}
  \caption{(Color online) Typical snapshots of the aqueous NaCl solution from (a) all-atom molecular dynamics (MD) simulations with explicit ions and (b) mesoscale charged dissipative particle dynamics (cDPD) simulations with semi-implicit ions.}
  \label{fig:1}
\end{figure}

To validate the closed-form expressions of fluctuation correlations we derived based on linearized fluctuating hydrodynamic and electrokinetic equations in Section~\ref{sec:theory}, we perform both all-atom MD simulations with explicit ions and mesoscale cDPD simulations with semi-implicit ions, as illustrated in figure~\ref{fig:1}. We first carry out all-atom MD simulations of an aqueous NaCl solution in the bulk, which consists of 400 Na$^+$ ions, 400 Cl$^-$ ions, and 64000 H$_2$O water molecules in a periodic cubic computational domain with box size $L = 12.8~{\rm nm}$. The mass density is $\rho^0 = 0.616~{\rm amu~\AA^{-3}}$, and the ion concentration is $c_{\pm} = 0.3168~{\rm M}$.
The SPC/E model~\citep{Berendsen1987} is used for water molecules. The ionic force field terms are adopted from published work by~\cite{Smith1994} and \cite{Yoshida2014}. The particle-mesh Ewald method~\citep{Eastwood1984} is used for computing electrostatic interactions with vacuum periodic boundary conditions and a direct space cutoff of $9.8~{\rm \AA}$.
The bond length of water molecules is constrained using the SHAKE algorithm~\citep{Ryckaert1977} to allow a time step of $1~{\rm fs}$ in the velocity-Verlet integrator~\citep{allen2017computer}.
A constant number-volume-temperature (NVT ensemble) system is simulated at $T = 300~{\rm K}$ with a Nos{\'e}-Hoover thermostat. The system is relaxed for $0.5~{\rm ns}$ to achieve a thermal equilibrium state, and then for up to $10~{\rm ns}$ for statistics and sampling.

In Section~\ref{sec:theory}, we assumed that the mutual diffusion is small compared to ionic self-diffusion, and thus it can be ignored in the derivation. To confirm this assumption, we compute both self-diffusion and mutual-diffusion coefficients of ionic species from the MD velocity correlation functions via the Green-Kubo relationship~\citep{1996Zhou,2004Wheeler}.
The computed self-diffusion coefficient is $1.3\times10^{-9}~{\rm m^2/s}$ for Na$^+$ and $2.1 \times 10^{-9}~{\rm m^2/s}$ for Cl$^{-}$, while their mutual-diffusion coefficient is $1.2 \times 10^{-10}~{\rm m^2/s}$. These results are consistent with previous simulation and experimental results for NaCl solutions~\citep{2004Wheeler}, and also confirm that the mutual-diffusion coefficient is much smaller than the self-diffusion coefficients in this electrolyte solution, and thus it can be ignored in the analytical derivations of current correlation functions presented in Section~\ref{sec:theory}.

Alternative to all-atom MD simulation, at the mesoscopic scale, a cDPD model is used to tackle the challenge of simulating coupled fluctuating hydrodynamics and electrostatics with long-range Coulomb interactions. The cDPD model is an extension of the classic DPD model to numerically solve the fluctuating hydrodynamics and electrokinetics equations in the Lagrangian framework. In classical DPD models, ions can be represented by explicit charged particles with electrostatic interactions between explicit ions. Groot~\citep{2003Groot} introduced a lattice to the DPD system to spread out the charges over the lattice nodes. Then, the long-range portion of the interaction potential was calculated by solving the Poisson equation on the grid based on a particle-particle particle-mesh (PPPM) algorithm~\citep{Hockney1988} by transferring quantities (charges and forces) from the particles to the mesh and vice versa. Because the mesh defines a coarse-grained length for electrostatic interactions, correlation effects on length scales shorter than the mesh size cannot be properly accounted for. Although the particle-to-mesh then mesh-to-particle mapping/redistribution can solve the Poisson equation for particle-based systems, its dependence on a grid may contradict the original motivation for using a Lagrangian method, and additional computational complexity and inefficiencies are introduced. To abandon grids and use an unifying Lagrangian description for mesoscopic electrokinetic phenomena, we developed the cDPD model~\citep{2016Deng} to solve the Poisson equation on moving cDPD particles rather than grids. More specifically, cDPD describes the solvent explicitly in a coarse-grained sense as cDPD particles, while the ion species are described semi-implicitly, i.e., using a Lagrangian description of ionic concentration fields, associated with each moving cDPD particle, as shown in Fig.~\ref{fig:1}(b), which provides a natural coupling between fluctuating electrostatics and hydrodynamics.

%%%
\begin{figure}
  \centering
  \includegraphics[width = 0.9\textwidth]{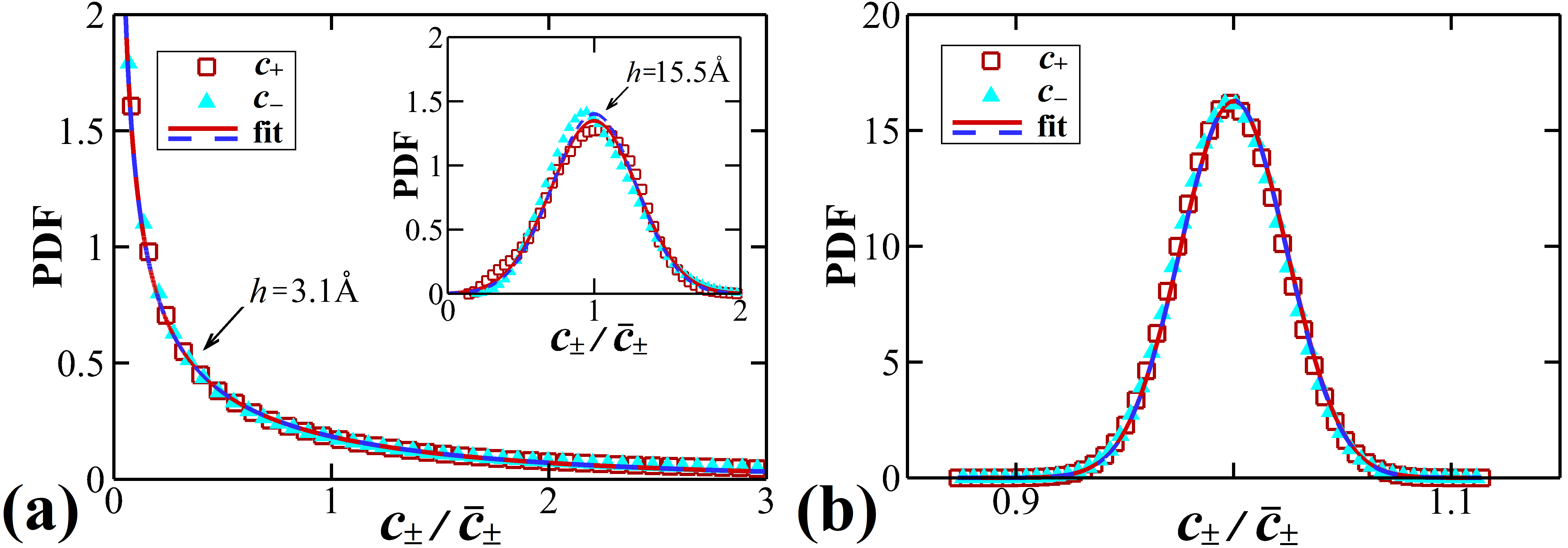}
  \caption{Local ionic concentration probability distribution functions from (a) all-atom MD and (b) mesoscale cDPD simulations. The symbols show simulation data while the lines show fits to Gamma (MD) and Gaussian (inset of (a), and cDPD) distributions.}
  \label{fig:2}
\end{figure}

The state vector of a cDPD particle can be written as $(\mathbf{r}, \mathbf{v}, c_{\alpha}, \phi)$, which is not only characterized by its position $\mathbf{r}$ and velocity $\mathbf{v}$ as in the classical DPD model, but also by ionic species concentration $c_{\alpha}$ (with $\alpha$ representing the $\alpha$th ion type) and electrostatic potential $\phi$ on the particle.\ A cDPD system is simulated in a cubic periodic computational box with length of $106.79~{\rm nm}$, where a cDPD particle is viewed as a coarse-grained fluid volume, which contains the solvent and other charged species. Exchange of the concentration flux of charged species occurs between neighboring cDPD particles, much like the momentum exchange in the classical DPD model. The governing equations of cDPD and model parameters are summarized in Appendix~\ref{sec:cdpd}. We will perform cDPD simulations of electrolyte solutions and compute current correlation functions of fluctuating electrokinetics to compare with theoretical predictions given in Section~\ref{sec:theory}.

%%%
\begin{figure}
\centering
\includegraphics[width = 0.4\textwidth]{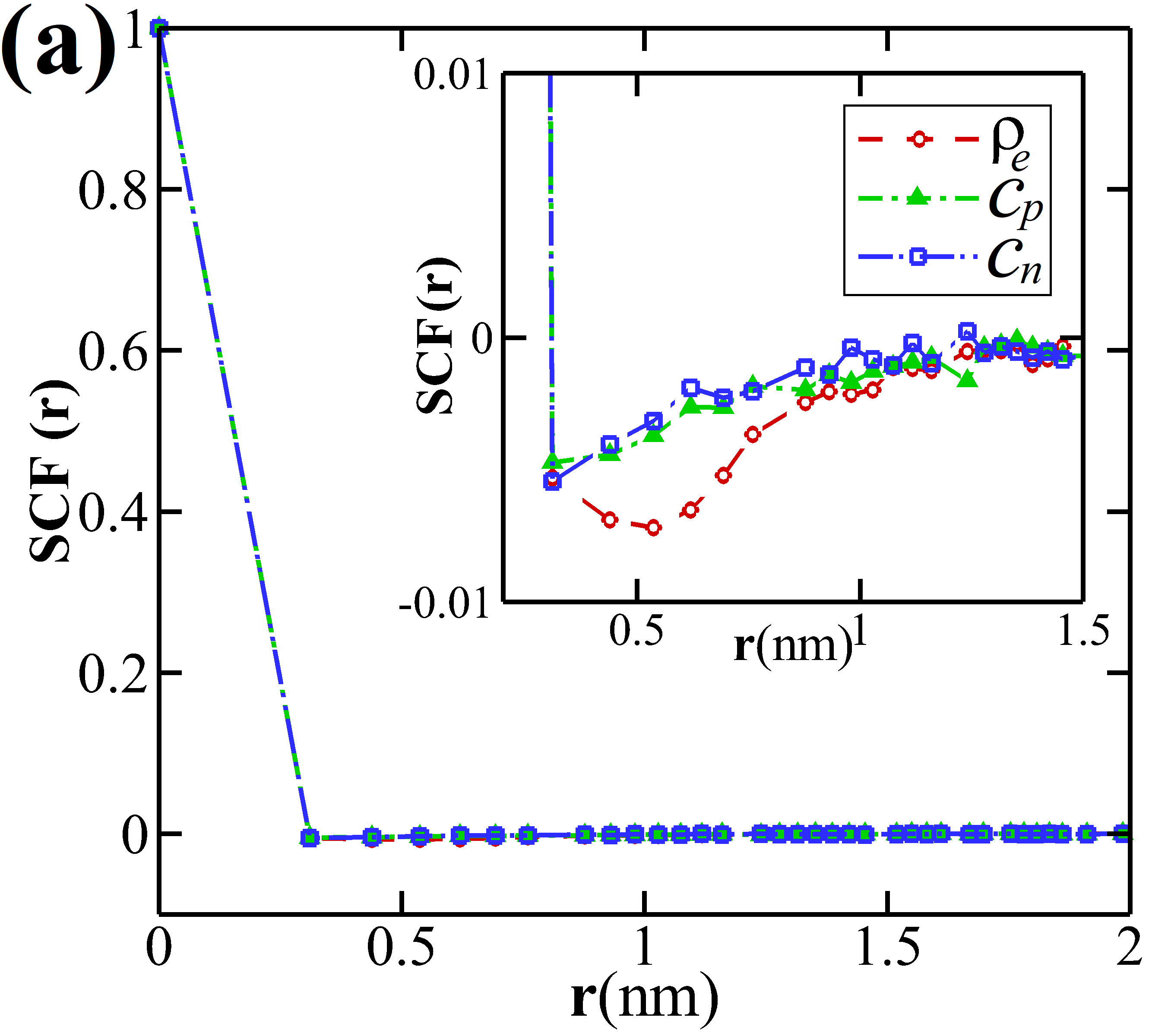}
\includegraphics[width = 0.4\textwidth]{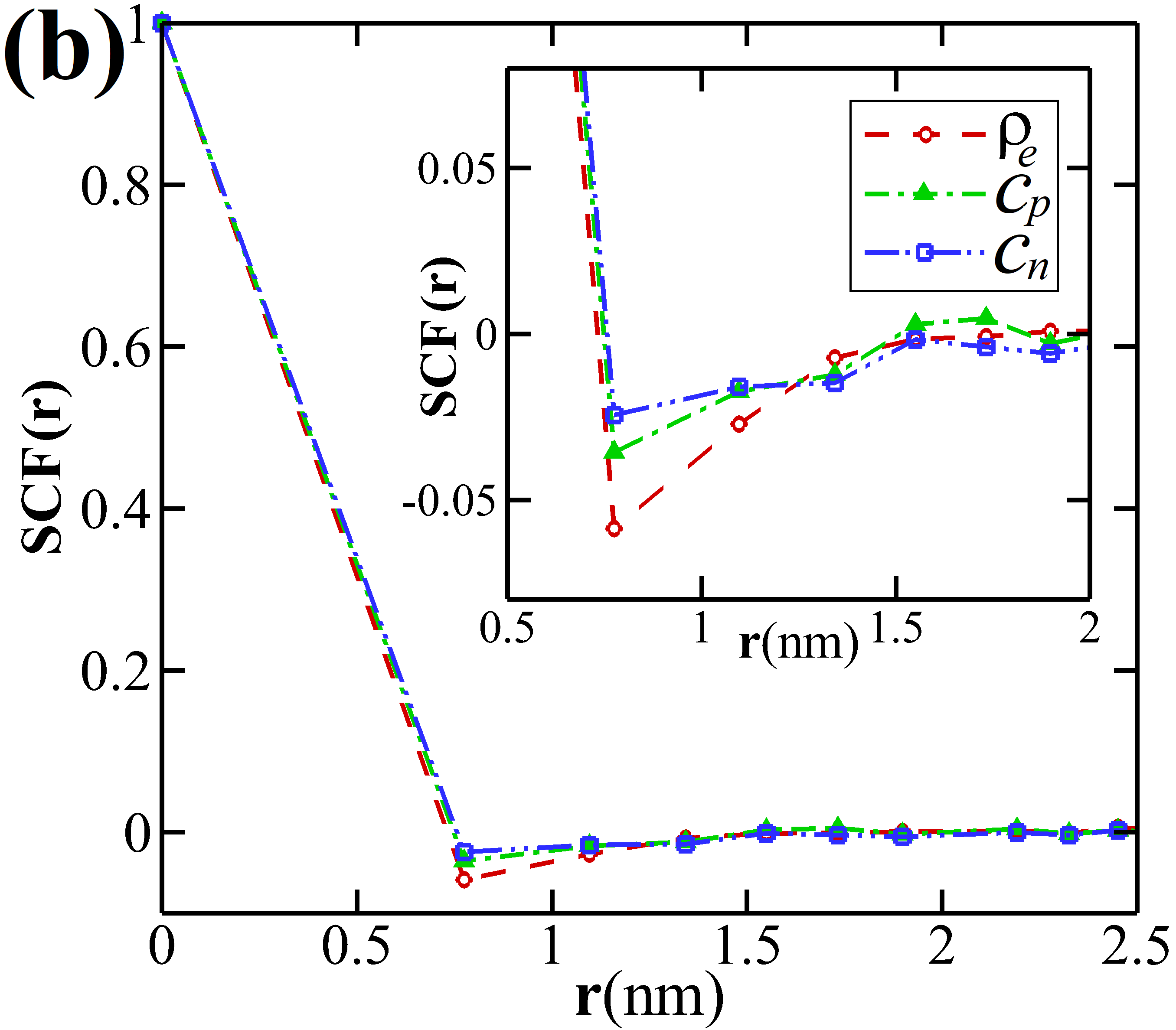}
\caption{Spatial correlation function (SCF) of charge density from full-atom MD simulations for grid distances of (a) $0.31~{\rm nm}$ and (b) $0.775~{\rm nm}$.}\vspace{-10pt}
\label{fig:3}
\end{figure}
%%%

We first examine the local fluctuations of mass density, momentum density, charge density and ionic concentration.
For each of these quantities, we construct the instantaneous function $n(\mathbf{r})$ on a grid $\mathbf{r}^{i,j,k}$ by spatial averaging using a step function $W(r, h)$ with $h$ being the grid size, i.e., $W(r, h) = 1$ for $|\mathbf{r}_i-\mathbf{r}|\le 0.5h$ and $W(r, h) = 0$ for $|\mathbf{r}_i-\mathbf{r}| > 0.5h$. Then the local quantities can be extracted by $n(\mathbf{r}) = \sum_i^N W(|\mathbf{r} - \mathbf{r}_i|, h) n_i$ with $h = 3.1~{\rm \AA}$ or $h = 7.75~{\rm \AA}$ for MD data and $h = 4.27~{\rm nm}$ for cDPD data.
According to central limit theorem, the local fluctuations in these quantities of interest should be Gaussian in the continuum regime. This is observed for mass density, momentum density, charge density from both MD and cDPD results.
However, the local concentration fluctuations for ionic species in the MD system follow Gamma distributions and converge to Gaussian with large spacing $h$ as shown in Fig.~\ref{fig:2}(a). Because the cDPD model assumes that the random fluxes of ionic concentration between neighboring cDPD particles can be modeled by Gaussian white noises, the MD results indicate that the cDPD model is valid for length scales above a few nanometers but has a lower bound for its length scale. Fig.~\ref{fig:2}(b) presents the probability distribution functions of local ionic concentration in cDPD simulations, which follow a Gaussian distribution as expected. 

\begin{figure}
	\centering
	\includegraphics[width = 0.5\textwidth]{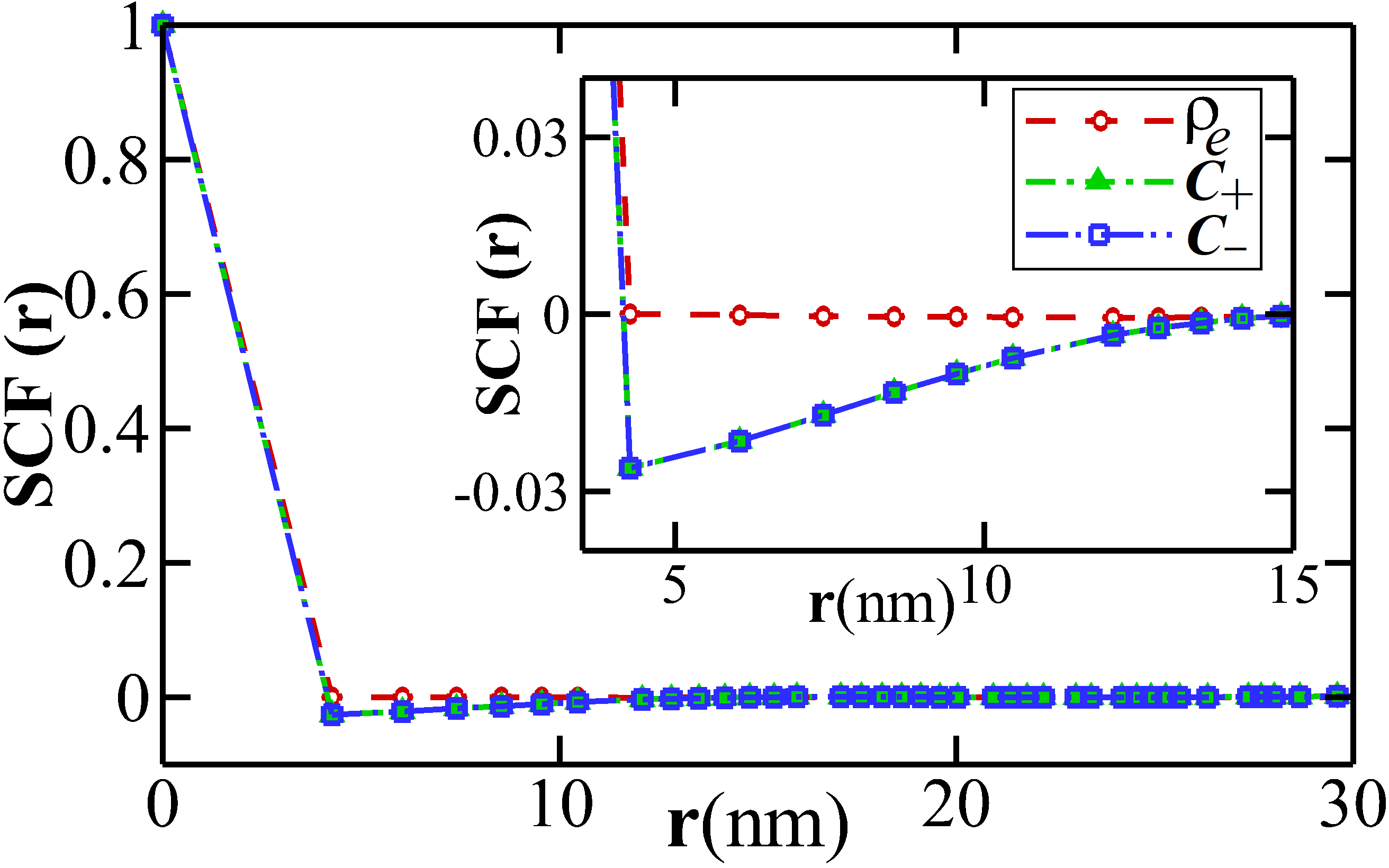}
	\caption{Spatial correlation function (SCF) of charge density from mesoscale cDPD simulations.}
	\label{fig:4}
\end{figure}

The spatial correlation function (SCF) for physical quantities of interest $n(\mathbf{r})$ are computed on the grid $\mathbf{r}^{i,j,k}$ according to
\vspace{-5pt}
\begin{equation}
	{\rm SCF}(\mathbf{r}) = \frac{1}{N(\mathbf{r})} \sum_{i=1}^{N(\mathbf{r})} \delta n(\mathbf{r}_i) \delta n(\mathbf{r}_i + \mathbf{r}),
\end{equation}
where $N(\mathbf{r})$ is defined as a normalization coefficient.
We have assumed that the spatial correlations are isotropic in weakly charged electrolyte bulk solutions.
We observe that the spatial correlations of mass density, momentum density, charge density and ion concentration are all very short ranged (approximately delta-correlated for large systems), which is in agreement with continuum fluctuating hydrodynamics results at equilibrium.
However, as shown in Fig.~\ref{fig:3}, there is small-scale anti-correlation structure for the charge densities obtained in the MD simulations. We also compute the spatial correlation function of charge density in mesoscale cDPD simulations, and find similar small-scale anti-correlation structures for both positive and negative ions, as shown in Fig.~\ref{fig:4}. These anti-correlation structures are observed on length scales comparable to the Bjerrum and Debye lengths for electrolyte solution systems, suggesting that it could be related to ionic screening effect. 
%%%

\begin{figure}
	\centering
	\includegraphics[width = 0.48\textwidth]{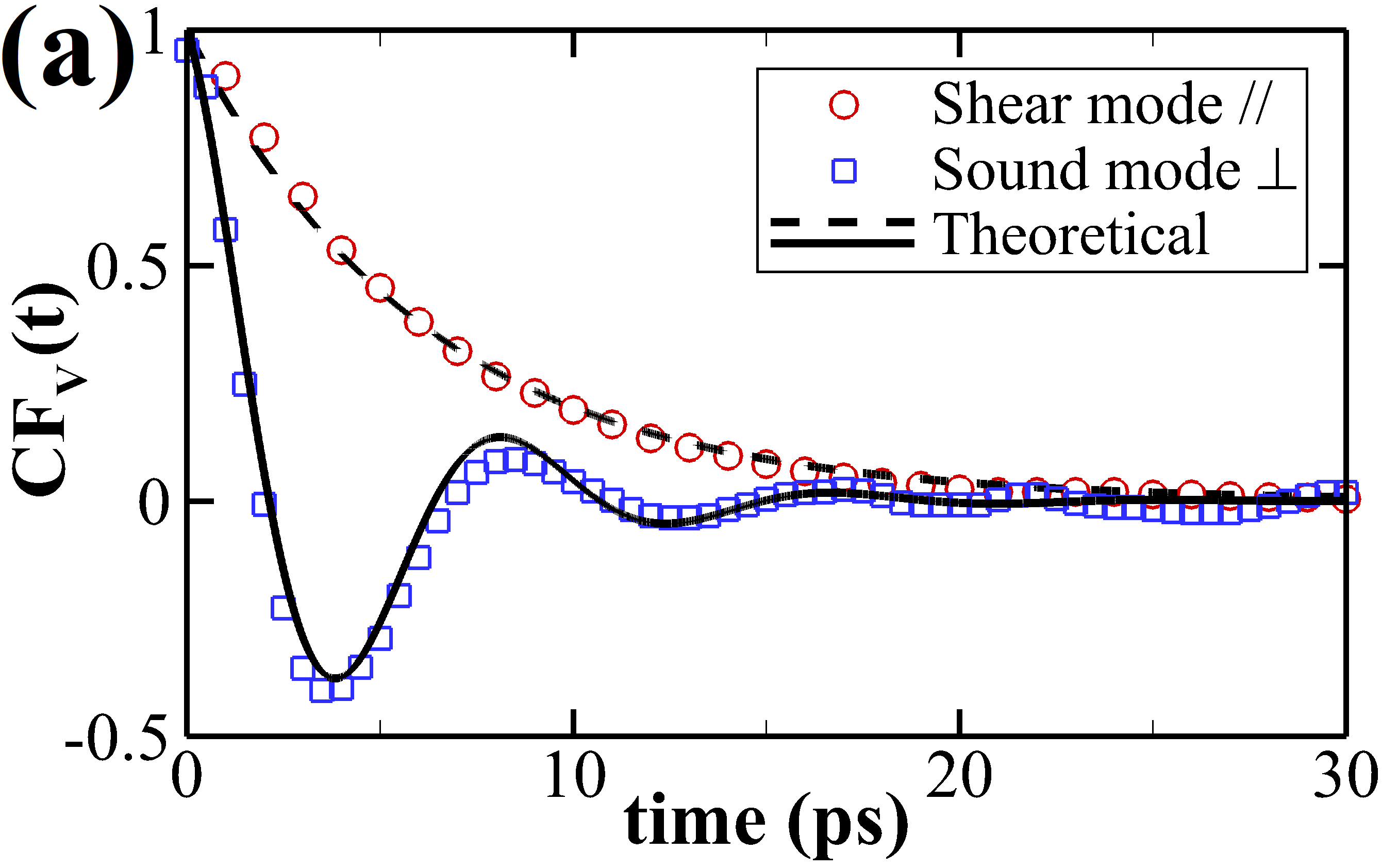}
	\includegraphics[width = 0.48\textwidth]{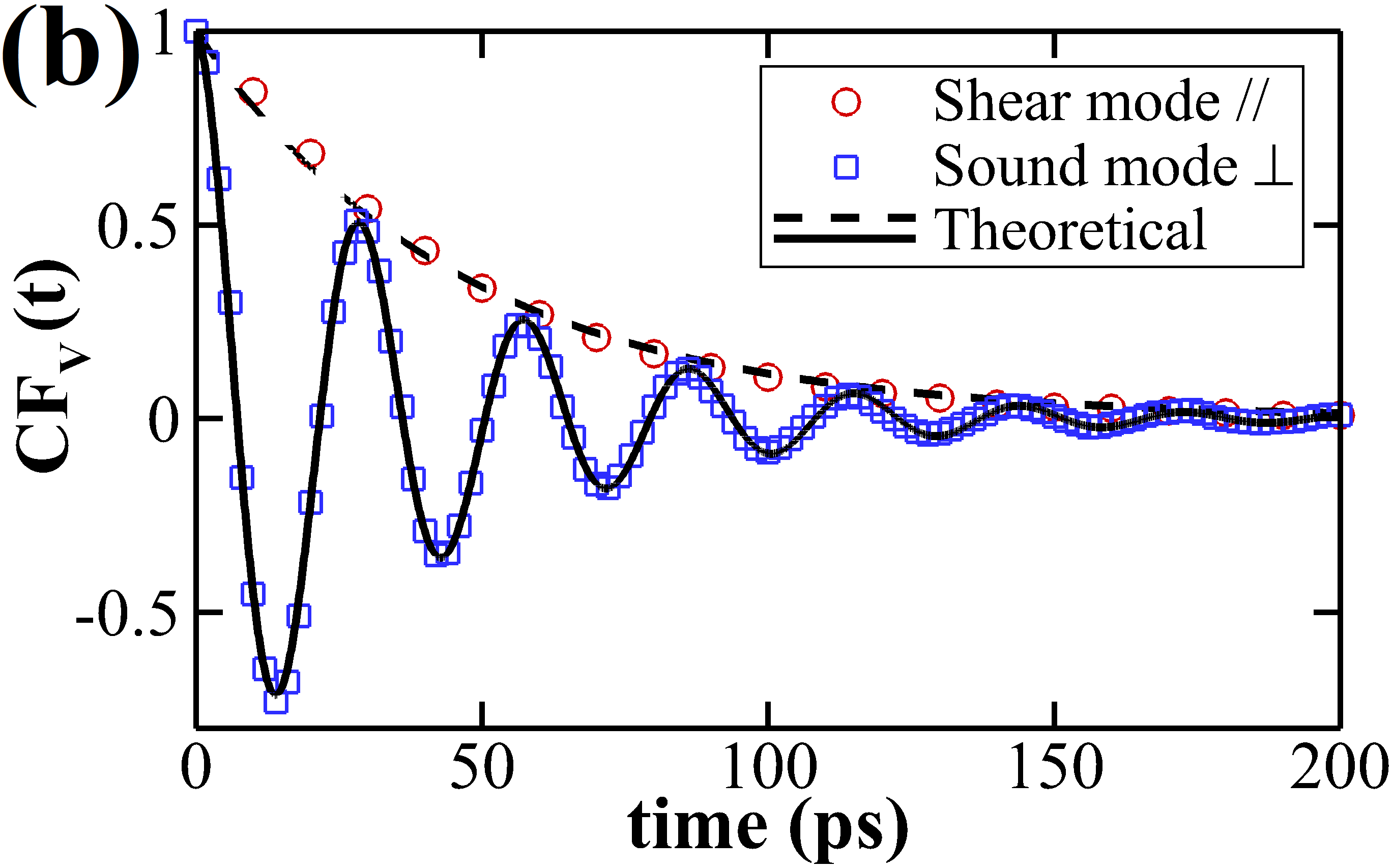}
	\caption{Temporal transverse (shear mode) and longitudinal (sound mode) velocity auto-correlation functions in Fourier space from (a) MD and (b) cDPD simulations represented by open symbols, with comparison against the theoretical predictions in the form of Eq.~(\ref{eq:ns_correlation}) in solid and dashed lines.}
	\label{fig:5}
\end{figure}

The temporal correlation function (TCF) between two physical quantities $u$ and $w$ in Fourier space is defined by
\begin{equation}
	{\rm TCF}_k(t)=\langle u(\mathbf{k},t) w(\mathbf{k}, 0)\rangle = \frac{1}{N(t)} \sum_{s=1}^{N(t)} \hat{u}(\mathbf{k}, t) \hat{w}(\mathbf{k}, 0),\nonumber
\end{equation}
where $\mathbf{k}$ is the wave vector in Fourier mode, $\displaystyle \hat{u}$, $\hat{w}$ are the Fourier components directly computed from particle trajectories as
\begin{equation}
	\hat{u}(\mathbf{k},t) = \frac{1}{N_P} \sum_{i=1}^{N_P} u_i(\mathbf{r}_i, t) \exp(-i \mathbf{k}\cdot\mathbf{r}_i(t)).\nonumber
\end{equation}
We use a wave length equals the computational box size $L$ by setting $k=2\pi/L$, which is $12.8~{\rm nm}$ for the MD system and $106.79~{\rm nm}$ for the cDPD system. 

Figs.~\ref{fig:5}(a) and~\ref{fig:5}(b) present the temporal velocity auto-correlation functions from MD and cDPD simulations, respectively. The theoretical predication based on Eq.~(\ref{eq:ns_correlation}) gives ${\rm CF}_v(t)=\exp(-0.1616t)$ (shear mode) and ${\rm CF}_v(t)=\exp(-0.2403t)\cos(0.7352t)$ (sound mode) for the MD system, and ${\rm CF}_v(t)=\exp(-0.0216t)$ (shear mode) and ${\rm CF}_v(t)=\exp(-0.0238t)\cos(0.2177t)$ (sound mode) for the cDPD system. It can be observed from Fig.~\ref{fig:5} that the computed results from both MD and DPD simulations are in agreement with the classical linearized theory of fluctuating hydrodynamics, i.e., the transverse autocorrelation function (shear mode) decays as $\exp(-\nu k^2 t)$, with $\nu$ the kinematic shear viscosity, while the longitudinal autocorrelation function (sound mode) decays as $\exp(-\Gamma_T k^2 t)\cos(c_s k t)$, with $\Gamma_T = 2\nu/3 + \zeta/2\rho$ the sound absorption coefficient and $c_s$ the isothermal sound speed.
We conclude that the fluctuating hydrodynamics model still follows the classical linearized theory, and is not explicitly affected by the electrolyte bulk solutions. In an electrolyte solution in thermodynamic equilibrium, the fluctuating hydrodynamics and electrokinetics are explicitly decoupled.

\begin{figure}
	\centering
	\includegraphics[width = 0.48\textwidth]{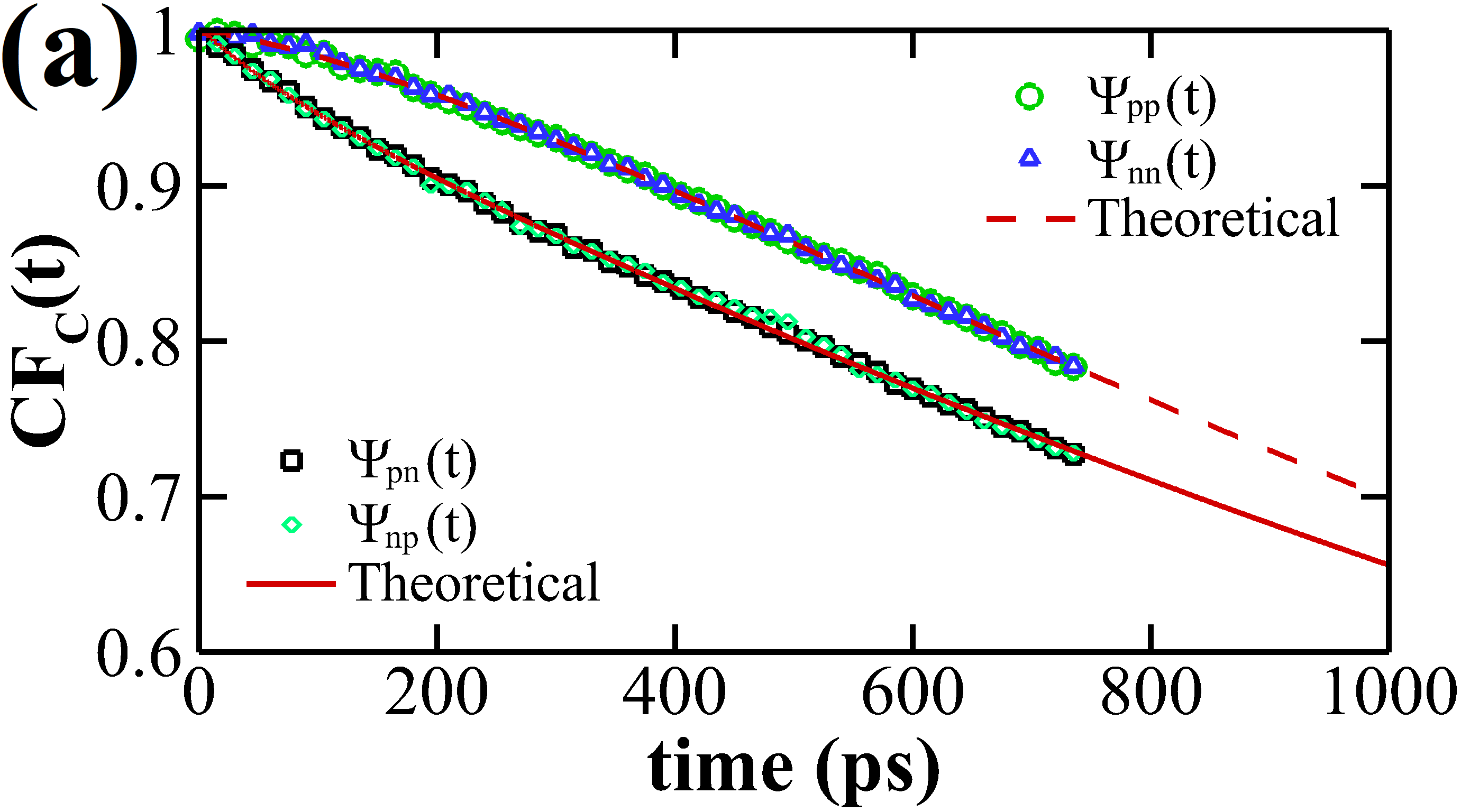}
	\includegraphics[width = 0.48\textwidth]{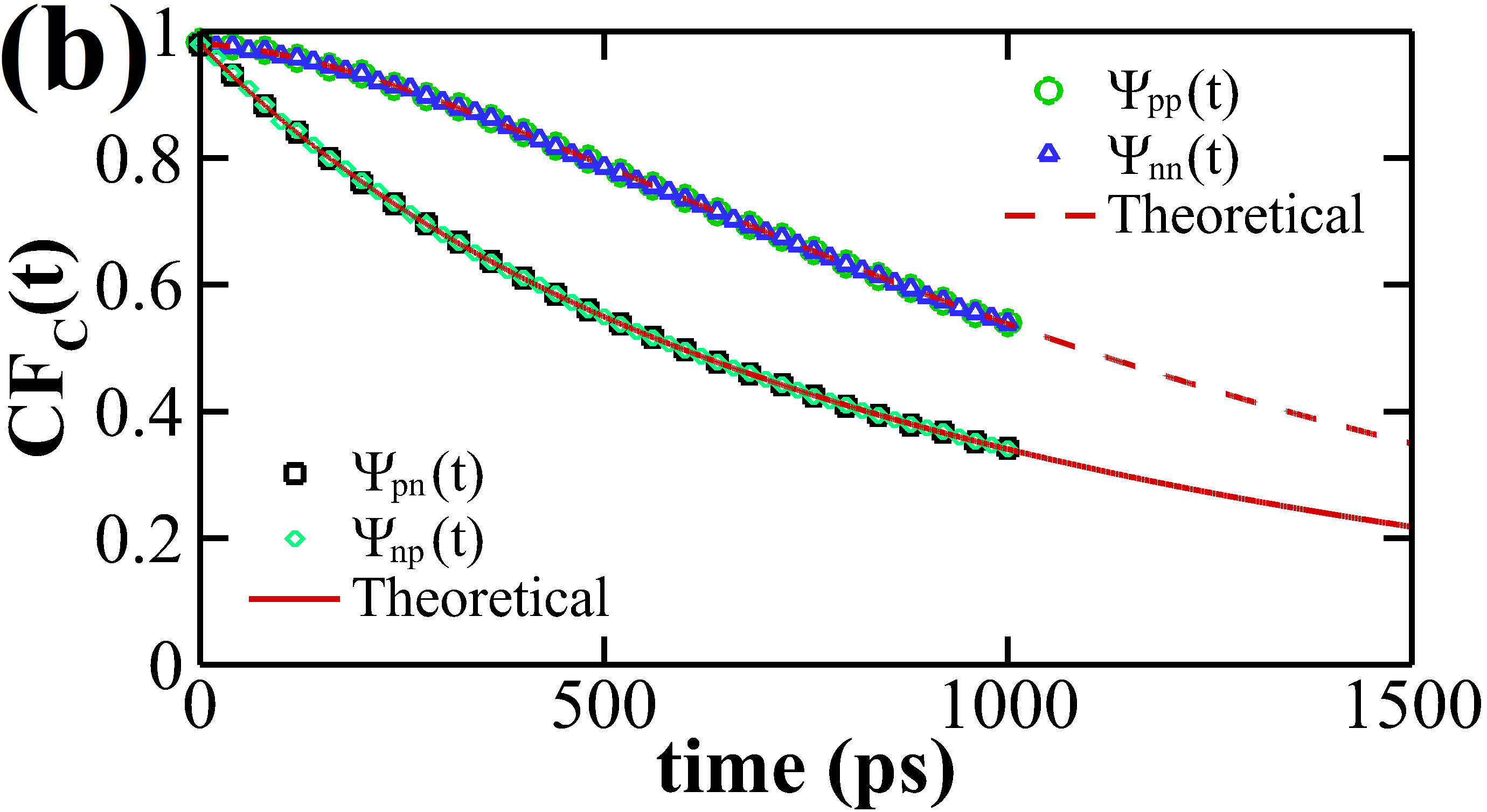}
	\caption{Temporal cation and anion concentration auto-correlation ($\Psi_{pp}$ and $\Psi_{nn}$) and cross-correlation ($\Psi_{pn}$ and $\Psi_{np}$) functions in Fourier space from (a) MD and (b) cDPD simulations represented by open symbols, with comparison against the predictions from linearized fluctuating hydrodynamics theory in the form of Eq.~(\ref{eq:TCF}) in solid and dashed lines.}
	\label{fig:6}
\end{figure}

Next, we compare the auto-correlation and cross-correlation functions of charge concentration with the linearized theory derived above, as shown in Fig.~\ref{fig:6}.
We fit the MD and DPD results with the linearized theory in the form of Eq.~(\ref{eq:TCF}) with proper structure factors $S_{p n}$ and $S_{n p }$, which are around $1.0$ and system dependent, i.e.,
\begin{subeqnarray}
    {\rm \Psi}_{pp}(t)&&={\rm \Psi}_{nn}(t)=-0.0490\exp(\lambda_1 t) + 1.0049\exp(\lambda_2 t),\nonumber\\
    {\rm \Psi}_{pn}(t)&&={\rm \Psi}_{np}(t)=~~0.0276\exp(\lambda_1 t) + 0.9724\exp(\lambda_2 t),\nonumber
\end{subeqnarray}
with $\lambda_1 = -7.4187\times10^{-3}$ and $\lambda_2 = - 3.9067\times10^{-4}$ for the MD system, and 
\begin{subeqnarray}
    {\rm \Psi}_{pp}(t)&&={\rm \Psi}_{nn}(t)=-0.2401\exp(\lambda_1 t) + 1.2401\exp(\lambda_2 t),\nonumber\\
    {\rm \Psi}_{pn}(t)&&={\rm \Psi}_{np}(t)=~~0.2458\exp(\lambda_1 t) + 0.7542\exp(\lambda_2 t),\nonumber
\end{subeqnarray}
with $\lambda_1 = -3.4446\times10^{-3}$ and $\lambda_2 = - 8.10\times10^{-4}$ for the cDPD system. 
We observe in Fig.~\ref{fig:6} that the computed results of temporal correlation functions from both MD and DPD simulations are in good agreement with the theoretical predictions derived from linearized theory of fluctuating hydrodynamics.
Compared to the temporal correlation functions of hydrodynamic fluctuations presented in Fig.~\ref{fig:5}, the temporal correlations of charge concentration fluctuations shown in Fig.~\ref{fig:6} decay extremely slowly over time. 
The long time behavior of both auto-correlation and cross-correlation follow the slow decay dominated by $\propto \exp(\lambda_2 t)$, which is significantly different from the decorrelation processes of hydrodynamic fluctuations.

\section{Summary}\label{sec:summary}
We have presented theoretical closed-form expressions for the fluctuations of electrolyte bulk solution close to thermodynamic equilibrium with an emphasis on mesoscopic spatiotemporal scales. In particular, we started with the Landau-Lifshitz theory and linearized the fluctuating hydrodynamic and electrokinetic equations to derive analytical solutions for current correlation functions using perturbation theory. To validate these theoretical expressions obtained based on the linearized theory, we performed numerical experiments of fluctuating hydrodynamics and electrokinetics for electrolyte solutions using both all-atom molecular dynamics (MD) and a mesoscale charged dissipative particle dynamics (cDPD) methods. 
We presented both MD and DPD simulation results of bulk electrolyte solutions in about 10~nm and 100~nm scales, which are directly compared with the predictions from the linearized continuum theory.

The current correlation functions computed from both MD and cDPD trajectories indicate that the temporal correlations of fluctuations from electrokinetics decay much more slowly than those from hydrodynamics, which agree well with the predictions made by the theoretical closed-form expressions of temporal correlation functions (hydrodynamics and charge) as well as the vast differences in decay time -- varying by a few orders of magnitude. 
In a bulk electrolyte solution close to thermodynamic equilibrium, the fluctuating hydrodynamics and electrokinetics are explicitly decoupled because of the zero mean velocity, and thus their behavior is not explicitly dependent on the electrolyte concentrations in the dilute regime.
At length scales above 10~nm, the results obtained from both MD and cDPD simulations are in good agreement with the continuum-limit linearized theories.
Spatial correlations of charge density demonstrate finite range and non-trivial structure at nanometer length scales but can be viewed as the delta function in the continuum limit. 
Simulation results also show that the fluctuations of local ionic concentration follow Gamma distribution at small length scales, while converge to Gaussian distribution in the continuum limit, which suggests the existence of a lower-bound of length scale for mesoscale models using Gaussian fluctuations for electrolyte solutions.

It is worth noting that the present work focused on dilute electrolyte solutions close to thermodynamic equilibrium, where the mutual-diffusion of ions is small compared to their self-diffusion and the fluctuating electrokinetics is decoupled from zero-mean hydrodynamic fluxes. When the mutual diffusion coefficients of ions become comparable to the self-diffusion terms in concentrated electrolyte solutions~\citep{2002Dufreche,2021Galindres}, the contribution of mutual-diffusion on ionic transport should be correctly considered in the fluctuating hydrodynamics and electrokinetics equations. It is also interesting to consider how the fluctuating hydrodynamics is coupled with mesoscale electrokinetics in shear flows, where the current correction functions can be affected by the coupling between fluctuating terms and advection processes~\citep{2018Bian}, leading to orientation-dependent decorrelation process of fluctuating variables in the fluid system. 

\section*{Acknowledgements}
This work was supported by an ARO/MURI Grant W911NF-15-1-0562 and the U.S. Army Research Laboratory under Cooperative Agreement No.\ W911NF-12-2-0023. F.~Tushar and Z.~Li acknowledge research support from Clemson University with generous allotment of compute time on the Palmetto cluster.

\section*{Declaration of Interests}
The authors report no conflict of interest.

\appendix
\section{Derivation of Mass-Momentum Correlations}\label{app:A}
The local hydrodynamic field can be expressed as the perturbations around the bulk state. We can define the instantaneous density, velocity and momentum as
\begin{subeqnarray}
        \rho(\mathbf{r},t) &&= \rho_0 + \delta \rho(\mathbf{r},t),\\
		\mathbf{u}(\mathbf{r},t) &&= \mathbf{u}_0 + \delta \mathbf{u}(\mathbf{r},t),\\
		\mathbf{g}(\mathbf{r},t) &&= \delta \mathbf{g} (\mathbf{r},t) = \rho\mathbf{u}(\mathbf{r},t).
\end{subeqnarray}
The linearized equations of fluctuating hydrodynamics are given by
\begin{subeqnarray}
        \frac{\partial [\delta \rho(\mathbf r,t)]}{\partial t} + \nabla \cdot \delta \mathbf g(\mathbf r,t) &&= 0, \\
        \frac{\partial [{\delta \mathbf g(\mathbf r,t)]}}{\partial t} + c_s^2 \nabla \delta \rho(\mathbf r,t) &&- \frac{\eta}{\rho_0} \nabla^2 \delta \mathbf g(\mathbf r,t)
         - \frac{\frac{\eta}{3} + \zeta}{\rho_0} \nabla (\nabla \cdot \delta\mathbf g(\mathbf r,t)) = 0.
\end{subeqnarray}
The above equations can be transformed into $k$-space by a spatial Fourier transform using the following function
\begin{equation}
    \hat{f_k}(t) = \int_V f(t) \exp(-i\mathbf k \cdot \mathbf r)d \mathbf r. \nonumber
\end{equation}
To make the derivation simple, we set the wave vector $\mathbf k = (k, 0, 0)$ as one-dimensional along an arbitrary $x$-direction, leading to
\begin{subeqnarray}
        &&\frac{\partial \, \widehat{\delta \rho}_k(t)}{\partial t} + i \mathbf k \cdot \widehat{\delta \mathbf g}_k (t) = 0, \\
        &&\frac{\partial \, \widehat{\delta \mathbf g}_k(t)}{\partial t} + ic_s^2 \mathbf k \widehat{\delta \rho}_k(t) + \frac{\eta}{\rho_0} k^2 \widehat{\delta \mathbf g}_k(t)
        + \frac{\frac{\eta}{3} + \zeta}{\rho_0} \, \mathbf k \mathbf k \, \widehat{\delta \mathbf g}_k(t) = 0. 
\end{subeqnarray}
We can divide the above equations into different components ($x, y$ and $z$ directions). The $x$-direction equations are
\begin{subeqnarray}
         &&\frac{\partial \, \widehat{\delta \rho}_k(t)}{\partial t} + ik \cdot \widehat{\delta g}_k^x (t) = 0, \\
         &&\frac{\partial \, \widehat{\delta g}_k^x(t)}{\partial t} + ic_s^2 k \widehat{\delta \rho}_k(t) + \frac{\eta}{\rho_0} k^2 \widehat{\delta g}_k^x(t)
         + \frac{\frac{\eta}{3} + \zeta}{\rho_0} \, k^2 \, \widehat{\delta g}_k^x(t) = 0.
\end{subeqnarray}
The $y$ and $z$ components are given by
\begin{subeqnarray}
        &&\frac{\partial \, \widehat{\delta g}_k^y(t)}{\partial t} + \frac{\eta}{\rho_0} k^2 \widehat{\delta g}_k^y(t) = 0, \\
        &&\frac{\partial \, \widehat{\delta g}_k^z(t)}{\partial t} + \frac{\eta}{\rho_0} k^2 \widehat{\delta g}_k^z(t) = 0.
\end{subeqnarray}
Let $\nu = {\eta}/{\rho_0}$ and $\nu_L = (\frac{4}{3}\eta + \zeta)/\rho_0$. The above questions can be rewritten into the following forms
\begin{subeqnarray}\label{eq:ODEs}
        &&\frac{\partial \, \widehat{\delta \rho}_k(t)}{\partial t} + ik \cdot \widehat{\delta g}_k^x (t) = 0, \\
        &&\frac{\partial \, \widehat{\delta g}_k^x(t)}{\partial t} + ic_s^2 k \widehat{\delta \rho}_k(t) + \nu_L \, k^2 \, \widehat{\delta g}_k^x(t) = 0, \\
        &&\frac{\partial \, \widehat{\delta g}_k^y(t)}{\partial t} + \nu k^2 \widehat{\delta g}_k^y(t) = 0, \\
        &&\frac{\partial \, \widehat{\delta g}_k^z(t)}{\partial t} + \nu k^2 \widehat{\delta g}_k^z(t) = 0.
\end{subeqnarray}
The last two equations are first-order linear ODEs, which are easily solved as
\begin{subeqnarray}
        \widehat{\delta g}_k^y(t) &&= \widehat{\delta g}_k^y(0) \exp (-\nu k^2 t), \\
        \widehat{\delta g}_k^z(t) &&= \widehat{\delta g}_k^z(0) \exp(-\nu k^2 t). 
\end{subeqnarray}
The first two equations of Eq.~(\ref{eq:ODEs}) are two coupled ODEs and can be written in a matrix form as
\begin{equation}\label{eq:matrix}
    \frac{d \mathbf{a}_k(t)}{dt} = \mathbf{H} \mathbf{a}_k(t),
\end{equation}
where 
\begin{subeqnarray}
		\mathbf a_k(t) = \left[ \begin{array}{c}
			\widehat{\delta \rho}_k(t)\\
			\\
			\widehat{\delta g}_k^x(t) \\
		\end{array} \right] 
		~~{\rm and ~~}
		\mathbf{H} = \left[ \begin{array}{cc}
			0 & -ik \\
			\, & \, \\
			-ic_s^2k & -\nu_Lk^2 \\
		\end{array} \right]. \nonumber
\end{subeqnarray}
The solutions to Eq.~(\ref{eq:matrix}) are determined by the eigenvalues of \textbf{\textit H}, which can be obtained by solving the equation
\begin{equation}
        {\rm det} (\mathbf{H} - \lambda \mathbf{I}) = 0. \nonumber
\end{equation}
The eigenvalues are given by
\begin{equation}
        \lambda_1 = -\Gamma_Tk^2 + is_Tk 
        ~~{\rm and~~}
        \lambda_2 = -\Gamma_Tk^2 - is_Tk, \nonumber
\end{equation}
where
\begin{equation}
      \Gamma_T = \frac{\nu_L}{2} ~~{\rm and ~~} s_T = \frac{\sqrt{4\,c_s^2 - \nu_L^2\,k^2}}{2}. \nonumber  
\end{equation}
$\Gamma_T$ is the sound absorption coefficient. Next, we consider an under-damped solution, where $s_T$ is real, that is, $k < 2c_s / \nu_L$. In particular, we consider the continuum limit, where $k \ll 2c_s / \nu_L$ so that $s_T \approx c_s$, then we arrive at the solutions as~\citep{2007Fabritiis,2018Bian}
\begin{subeqnarray}
        \widehat{\delta \rho}_k(t) &&= e^{-\Gamma_T k^2 t}\left[ \cos(c_skt) \widehat{\delta \rho}_k(0) - \frac{i}{c_s} \sin(c_skt) \widehat{\delta g}_k^x(0) \right], \\
        \widehat{\delta g}_k^x(t) &&= e^{-\Gamma_T k^2 t}\left[ \cos(c_skt) \widehat{\delta g}_k^x(0) - ic_s \sin(c_skt) \widehat{\delta \rho}_k(0) \right]. 
\end{subeqnarray}
Therefore, the normalized temporal correlation functions (TCF) of the mass-momentum fluctuations in the $k$-space are given by
\begin{subeqnarray}
       \frac{\langle \widehat{\delta \rho}_k(t) \widehat{\delta \rho}_k(0) \rangle}{\langle \widehat{\delta \rho}_k(0) \widehat{\delta \rho}_{k}(0) \rangle} &&= e^{-\Gamma_T k^2 t} \cos (c_s kt),\\
       \frac{\langle \widehat{\delta g}_k^x(t) \widehat{\delta g}_k^x(0) \rangle}{\langle \widehat{\delta g}_k^x(0) \widehat{\delta g}_k^x(0) \rangle} &&= e^{-\Gamma_T k^2 t} \cos (c_s kt),\\
       \frac{\langle \widehat{\delta \rho}_k(t) i\widehat{\delta g}_k^x(0) \rangle}{\langle \widehat{\delta \rho}_k(0) i\widehat{\delta g}_k^x(0) \rangle} &&= e^{-\Gamma_T k^2 t} \sin (c_s kt),
\end{subeqnarray}
where we have assumed that the initial cross correlation $\langle \widehat{\delta \rho}_k(0) \widehat{\delta g}_k^x(0) \rangle = 0$.

\section{Charged Dissipative Particle Dynamics (cDPD)}\label{sec:cdpd}
We consider a constant number-volume-temperature system (NVT ensemble) consisting of $N$ cDPD particles, with the state of each cDPD  particle defined by its position $\mathbf{r}$, velocity $\mathbf{v}$, and ionic concentration $c_{\alpha}$ (with $\alpha$ represents the $\alpha$-th ion species).
The time evolution of  the $i$-th particle state with unit mass is governed by Newton's law and transport equation~\citep{2016Deng}:
\begin{subeqnarray}\label{newton_eq}
	&&\frac{d^2 \mathbf{r}_i}{dt^2} = \frac{d\mathbf{v}_i}{dt} = \mathbf{F}_i = \sum_{i \neq j}(\mathbf{F}^C_{ij}+\mathbf{F}^D_{ij}+\mathbf{F}^R_{ij}+\mathbf{F}^E_{ij}), \\
	&&\frac{d c_{\alpha i}}{dt} = q_{\alpha i} = \sum_{i \neq j}(q^D_{\alpha ij} + q^E_{\alpha ij} + q^R_{\alpha ij}),
\end{subeqnarray}
where $\mathbf{F}_i$ denotes the total force exerted on the $i$-th particle, which consists of the conservative, dissipative and random forces.
Additionally, the electrostatic force $\mathbf{F}_i^E$ is introduced to couple the hydrodynamics and electrokinetics within the DPD framework. In particular,
\begin{subeqnarray}
		\mathbf{F}^C_{ij} &&= a_{ij}\omega_C(r_{ij})\hat{\mathbf{r}}_{ij}, \\
		\mathbf{F}^D_{ij} &&= - \gamma_{ij} \omega_D(r_{ij})(\hat{\mathbf{r}}_{ij}\cdot\mathbf{v}_{ij}))\hat{\mathbf{r}}_{ij},\\
		\mathbf{F}^R_{ij} &&= \sigma_{ij}\omega_R(r_{ij})\theta_{ij} \delta t^{-1/2} \hat{\mathbf{r}}_{ij},\\
		\mathbf{F}^E_{ij} &&= \lambda_{ij} (\sum_\alpha z_\alpha c_{\alpha i})\mathbf{E}_{ij},
\end{subeqnarray}
where $r_{ij}=|\mathbf{r}_{ij}|=|\mathbf{r}_i-\mathbf{r}_j|$, $\hat{\mathbf{r}}_{ij}=\mathbf{r}_{ij}/r_{ij}$ and $\mathbf{v}_{ij}=\mathbf{v}_i-\mathbf{v}_j$.
The conservative, dissipative, and random forces are pairwise forces with weighting functions $\omega_{C}(r_{ij})$, $\omega_D(r_{ij})$, $\omega_R(r_{ij})$, and corresponding strength $a_{ij}$, $\gamma_{ij}$, $\sigma_{ij}$, respectively.
$\theta_{ij}$ are symmetric Gaussian random variables with zero means and unit variances; these variables are independent for different pairs of particles at different times; $\theta_{ij} = \theta_{ji}$ is enforced to satisfy momentum conservation.
The dissipative and random forces together act as a thermostat with their coefficients and weighting functions satisfying the fluctuation-dissipation theorem (FDT)~\citep{Espanol1995}
\begin{equation}
	\sigma_{ij}^2=2k_BT\gamma_{ij},  \quad \quad \omega_D(r_{ij})=\omega_R^2(r_{ij}),
\end{equation}
where $k_B$ is the Boltzmann constant and $T$ the temperature.
The coupling parameter $\lambda_{ij}$ in electrostatic force is introduced by rescaling the PNP equations with DPD units, which is linearly related to the macroscopic dimensionless coupling parameter $\displaystyle \Lambda = c_0^* \cdot k_BT \tau^2/({\rho_0 r_0^5})$ with $c_0^* = c_0 r_0^3$, the reference concentration in DPD units (usually $c_0$ as
bulk concentration, $r_0$ the unit DPD length).
$\mathbf{E}_{ij}$ is the relative electric fields difference between particle $i$ and $j$, which is determined by the electrostatic potential field $\phi$
\begin{equation}
	\mathbf{E}_{ij} = \Big((\phi_i - \phi_j)\Big) \omega_E(r_{ij})\mathbf{r}_{ij}
\end{equation}
with ${\omega}_E(r)$ weighting function.
It is very important to note that the electrostatic forces here are not pairwise additive; i.e., $\mathbf{F}^E_{ij} \neq \mathbf{F}^E_{ji}$, however, $\sum_{i,j} \mathbf{F}^E_{ij} = 0$ to guarantee the global momentum conservation when there is no external electrostatic field.
In the cDPD framework, the ionic concentration (rescaled by the reference concentration $c_0$) evolution is driven by three pairwise flux terms; i.e., the Fickian flux $q_{\alpha ij}^{D}$, electrostatic flux $q_{\alpha ij}^{E}$ and random flux $q_{\alpha ij}^{R}$ induced by concentration gradient, electrostatic potential gradient and thermal fluctuations, respectively.
Specifically,
\begin{subeqnarray}
		q_{\alpha ij}^{D} &&= -\kappa_{\alpha ij}(c_{\alpha i}-c_{\alpha j})\omega_{DC}(r_{ij}) \\
		q_{\alpha ij}^{E} &&= -\frac{1}{2}\kappa_{\alpha ij}z_\alpha (c_{\alpha i} + c_{\alpha j})(\phi_i - \phi_j)\omega_{DC}(r_{ij}) \\
		q_{\alpha ij}^{R} &&= \xi_{\alpha ij} \omega_{RC} (r_{ij}) \theta_{ij} \delta t^{-1/2}
\end{subeqnarray}
where $\kappa_{ij}$ the diffusion coefficients, and $\omega_{DC}(r_{ij})$ and $\omega_{RC}$ weighting functions.
The coefficients and weighting functions of the random flux are determined via the generalized FDT as
\begin{equation}\label{FDC}
	{\xi_{\alpha ij}}^2 = \frac{\kappa_{\alpha ij}}{c_0^*}(c_{\alpha i}+c_{\alpha j}), \quad \quad \omega_{DC}(r)=\omega_{RC}^2(r)
\end{equation}
The electrostatic potential $\phi$ on each cDPD particle is determined by solving a modified Poisson equation at every DPD timestep.
In cDPD, we consider the dimensionless modified Poisson equation rescaled by DPD units
\begin{equation}
	\nabla \cdot \bigg(\epsilon (\mathbf{r}) \nabla (\phi(\mathbf{r})\bigg) = -\Gamma \rho_e(\mathbf{r})
\end{equation}
with $\Gamma=e^2c_0^*r_0^2/\epsilon_0 k_BT$ and $\rho_e = \sum_\alpha z_\alpha c_\alpha$ the charge density.
The electrostatic potential $\phi_i$ on the $i$th particle is obtained together via a SOR iteration scheme as
\begin{equation}\label{equ:SOR}
	\phi_i^{k} = \phi_i^{k-1} + \vartheta\left[\sum_\alpha^m \Gamma z_\alpha c_\alpha - \sum_{j \neq i}\bar{\epsilon}_{ij} \phi_{ij}^{k} \omega_{\phi}(r_{ij}) \right]
\end{equation}
where $\omega_{\phi}(r)$ the weight function, $k$ represents an iteration step, and $\vartheta$ the relaxation factor; $\epsilon_i$ and $\epsilon_j$ are the permittivity of the $i$th and $j$th cDPD particles, respectively.
$\bar{\epsilon}_{ij}=(\epsilon_i + \epsilon_j)/2$, where $\epsilon_i$ and $\epsilon_j$ can be different values to model mixtures of heterogeneous solvents.
The initial guesses of $\phi_i^{k-1}$ take the value of $\phi_i$ from the previous time step.
The iteration stops when the absolute differences $|\phi_i^{k}-\phi_i^{k-1}|$  are smaller than a tolerance; i.e., $10^{-3}$ for all cDPD particles.
The relaxation factor $\vartheta$ is adaptively selected during the iteration to optimize for faster convergence.
For more details on the derivations leading to Eqs.~(\ref{newton_eq})-(\ref{equ:SOR}), we refer interested readers to a previous work~\cite{2016Deng}.

Throughout this paper, the cDPD parameters are selected as $\rho = 4.0$, $k_BT = 1.0$, $a = 18.75$, $\gamma = 4.5$, $\kappa = 0.25$, $\lambda = 0.1438$, $\xi = 6.0$, $c_0^* = 26.224$,  $r_c = 1.5$, $r_{cc} = 1.0$, $r_{ec} = 2.5$. The weighting functions are chosen as $\omega_C(r) = (1-r/r_c)$, $\omega_D(r) = \omega_R^2(r) = (1-r/r_c)$, $\omega_{qD}(r) = \omega_{qR}^2(r) = (1-r/r_{cc})^2$, $\omega_\phi(r) = (1-r/r_{ec})^2$ and $\omega_E(r) = 0.5(1-r/r_{ec})^2 r$. The basic DPD units according to these parameters are $r_0=21.36$~nm for the length unit, $\tau=3.28$~ns for the time unit, $k_BT=4.14\times10^{-21}$~J for the energy units and $c_0=4.08\times10^{-3}$~M for the concentration unit.

\bibliographystyle{jfm}
% Note the spaces between the initials
\bibliography{references}

\end{document}